\documentclass[debug,overfull,figures,info,cite]{epl}

\usepackage{amsmath}
\usepackage{graphicx}
\usepackage{amssymb}
\usepackage{bm}

\newcommand{\be}{\begin{equation}}
\newcommand{\ee}{\end{equation}}
\newcommand{\bea}{\begin{eqnarray}}
\newcommand{\eea}{\end{eqnarray}}

\title{One-Electron Singular Branch Lines of the Hubbard Chain}
\author{J. M. P. Carmelo\inst{1}\thanks{E-mail: \email{carmelo@fisica.uminho.pt}}
\and K. Penc\inst{2} \and L. M. Martelo\inst{1,3} \and P. D.
Sacramento\inst{4} \and J. M. B. Lopes dos Santos\inst{5} \and R.
Claessen\inst{6} \and M. Sing \inst{6} \and U.
Schwingenschl\"ogl\inst{6}} \shortauthor{J. M. P. Carmelo \etal}
\institute{ \inst{1} GCEP-Center of Physics, U. Minho, Campus
Gualtar,
P-4710-057 Braga, Portugal\\
\inst{2} Res. Inst. for Solid State Physics
and Optics, H-1525 Budapest, P.O.B. 49, Hungary\\
\inst{3} Physics Department, Engineering Faculty of U.
Porto, P-4200-465 Porto, Portugal\\
\inst{4} CFIF - Instituto Superior T\'ecnico,
Av. Rovisco Pais, 1049-001 Lisboa, Portugal\\
\inst{5} CFP and Departamento de F\'{\i}sica FC U. Porto,
P-4169-007 Porto, Portugal \\
\inst{6} Experimentalphysik II, Universit{\"a}t Augsburg, D-86135
Augsburg, Germany}

\begin{document}

\maketitle

\begin{abstract}
The momentum and energy dependence of the weight distribution in
the vicinity of the one-electron spectral-function singular branch
lines of the 1D Hubbard model is studied for all values of the
electronic density and on-site repulsion $U$. To achieve this goal
we use the recently introduced pseudofermion dynamical theory. Our
predictions agree quantitatively for the whole momentum and energy
bandwidth with the peak dispersions observed by angle-resolved
photoelectron spectroscopy in the quasi-1D organic conductor
TTF-TCNQ.
\end{abstract}

The finite-energy spectral dispersions recently observed in
quasi-one-dimensional (1D) metals by angle-resolved photoelectron
spectroscopy (ARPES) reveal significant discrepancies from the
conventional band-structure description \cite{Ralph,Zwick98}. The
study of the microscopic mechanisms behind these unusual
finite-energy spectral properties remains until now an interesting
open problem. There is some evidence that the correlation effects
described by the 1D Hubbard model might contain such finite-energy
mechanisms \cite{Ralph,Zwick98}. However, for finite values of the
on-site repulsion $U$ very little is known about its finite-energy
spectral properties, in contrast to simpler models \cite{Arikawa}.
Bosonization \cite{Schulz} and conformal-field theory \cite{CFT}
do not apply at finite energy. For $U\rightarrow\infty$ the method
of Ref. \cite{Penc} provides valuable qualitative information, yet
a quantitative description of the finite-energy spectral
properties of quasi-1D metals requires the solution of the problem
for finite values of $U$. The method of Ref. \cite{Sorella} refers
to features of the insulator phase. For $U\approx 4t$, where $t$
is the transfer integral, there are numerical results for the
one-electron spectral function \cite{Senechal} which,
unfortunately, provide very little information about the
microscopic mechanisms behind the finite-energy spectral
properties. Recent preliminary results obtained by use of the
finite-energy holon and spinon description introduced in Refs.
\cite{I,II,IIIb} predict separate one-electron charge and spin
spectral branch lines \cite{Ralph}. For the electron-removal
spectral function these lines show quantitative agreement with the
peak dispersions observed by ARPES in the quasi-1D organic
conductor TTF-TCNQ \cite{Ralph}. However, these preliminary
studies provide no information about the momentum and energy
dependence of the weight distribution in the vicinity of the
charge and spin branch lines and do not describe the TTF
dispersion. The main goal of this paper is the evaluation of such
a dependence for all values of $U$ and electronic density. In
order to solve this complex many-electron problem, we use the
pseudofermion dynamical theory recently introduced in Ref.
\cite{theory}.

The model reads $\hat{H}=-t\sum_{j,\,\sigma}[c_{j,\,\sigma}^{\dag}
c_{j+1,\,\sigma} + h.
c.]+U\sum_{j}\hat{n}_{j,\uparrow}\hat{n}_{j,\downarrow}$ where
$c_{j,\,\sigma}^{\dagger}$ and $c_{j,\,\sigma}$ are spin $\sigma
=\uparrow ,\downarrow$ electron operators at site $j=1,...,N_a$
and
$\hat{n}_{j,\,\sigma}=c_{j,\,\sigma}^{\dagger}\,c_{j,\,\sigma}$.
The low-energy spectral properties of TTF-TCNQ involve inter-chain
hopping and electron-phonon interactions. Thus, our results are to
be applied above the energies of these processes. We consider an
electronic density $n=N/N_a$ in the range $0<n< 1$ and zero
magnetization where $N$ is the electron number. The Fermi momentum
is $k_F=\pi n/2$ and the electronic charge reads $-e$. The
one-electron spectral function $B^{l} (k,\,\omega)$ such that
$l=-1$ (and $l=+1$) for electron removal (and addition) reads,
$B^{-1} (k,\,\omega)=\sum_{\sigma,\,\gamma}
\vert\langle\gamma\vert\, c_{k,\,\sigma} \vert \,GS\rangle\vert^2
\,\delta (\omega +\Delta E_{\gamma}^{N-1})$ and $B^{+1}
(k,\,\omega)=\sum_{\sigma,\,\gamma'} \vert\langle\gamma'\vert\,
c^{\dagger}_{k,\,\sigma} \vert \,GS\rangle\vert^2 \,\delta (\omega
-\Delta E_{\gamma'}^{N+1})$. Here $c_{k,\,\sigma}$ and
$c^{\dagger}_{k,\,\sigma}$ are electron operators of momentum $k$
and $\vert GS\rangle$ denotes the initial $N$-electron ground
state. The $\gamma$ and $\gamma'$ summations run over the $N-1$
and $N+1$-electron excited states, respectively, and $\Delta
E_{\gamma}^{N-1}$ and $\Delta E_{\gamma'}^{N+1}$ are the
corresponding excitation energies.

The pseudofermion dynamical theory reveals and characterizes the
dominant microscopic processes which generate over 99\% of the
electronic weight of the functions $B^{l} (k,\,\omega)$, and also
describes all other processes \cite{theory}. The weight
distribution in the vicinity of the branch lines we consider below
is fully controlled by such dominant processes. Those involve the
$-1/2$ Yang holons and the $c0$,\,\, $s1$, and $c1$ pseudofermions
studied in Ref. \cite{IIIb}. For simplicity, we denote the first
two objects by $c$ and $s$ pseudofermions, respectively. The $c$
pseudofermion carries charge $-e$ and has no spin and the $s$
pseudofermion is a spin-zero two-spinon composite object and has
no charge. The $c1$ pseudofermion is a $\eta$-spin-zero two-holon
composite object, carries charge $-2e$, and has zero spin. The
$-1/2$ Yang holon has $\eta$-spin $1/2$, $\eta$-spin projection
$-1/2$, charge $-2e$, and zero spin. The $c$, $s$, and $c1$
pseudofermions carry momentum  ${\bar{q}} = q + Q_{\alpha}
(q)/N_a$, where in our case $\alpha =c,s,c1$. Here $q$ is the {\it
bare-momentum} and $Q_{\alpha}(q)/N_a$ is a momentum functional
defined in Ref. \cite{IIIb}, whose expression involves the
two-pseudofermion phase shifts $\Phi_{\alpha\,\alpha'}(q,q')$
defined in the same reference, where $\alpha,\,\alpha'=c,s,c1$.
Following the one-to-one correspondence between the momentum
${\bar{q}}$ and bare-momentum $q$, one can either label a $\alpha$
pseudofermion by ${\bar{q}}$ or $q$. Here we use the bare-momentum
$q$. The above pseudofermions have energy bands $\epsilon_{c}
(q)$, $\epsilon_{s} (q)$, and $\epsilon_{c1} (q) = E_u
+\epsilon_{c1}^0 (q)$ such that $\vert\,q\vert\leq\pi$,
$\vert\,q\vert\leq k_F$, and $\vert\,q\vert\leq [\pi -2k_F]$,
respectively. These bands are studied and plotted in Ref.
\cite{II}. $E_u$ defines the lower-limit of the {\it upper-Hubbard
band} (UHB) \cite{I} and equals the energy required for creation
of a $-1/2$ Yang holon, which is a dispersion-less object. It is
such that $E_u = 4t\cos (\pi n/2)$ for $U/t\rightarrow 0$, $E_u=U+
4t\cos (\pi n)$ for $U>>t$, $E_u = U + 4t$ for $n\rightarrow 0$,
and as $n\rightarrow 1$ $E_u$ approaches the value of the
Mott-Hubbard gap \cite{II}. In the ground state there are no
$-1/2$ Yang holons, the $c1$ and $s$ pseudofermions bands are
empty and filled, respectively, and the $c$ pseudofermions occupy
$0\leq\vert\,q\vert\leq 2k_F$ (leaving $2k_F<\vert\,q\vert\leq\pi$
empty).

The ground state and excited states can be expressed in terms of
occupancy configurations of the above quantum objects. For
electron removal, the dominant processes involve creation of one
hole both in $\epsilon_c (q)$ and $\epsilon_{s} (q)$. For electron
addition, these dominant processes lead to two structures: A {\it
lower-Hubbard band} (LHB) generated by creation of one particle in
$\epsilon_c (q)$ and one hole in $\epsilon_{s} (q)$; A UHB
generated by creation of one hole both in $\epsilon_c (q)$ and
$\epsilon_{s} (q)$ and either one particle in $\epsilon^0_{c1}
(q)$ for $n<1$ or one $-1/2$ Yang holon for $n\rightarrow 1$.
According to the pseudofermion dynamical theory of Ref.
\cite{theory}, both the one-electron spectral-weight singularities
and edges are located on pseudofermion branch lines. Such lines
are generated by processes where a specific pseudofermion is
created or annihilated for the available values of bare-momentum
$q$ and the remaining quantum objects are created or annihilated
at their {\it Fermi points}. The weight shape of the singular (and
edge) branch lines is controlled by negative (and positive)
exponents smaller than zero (and one). The electron removal
($\omega <0$) and LHB addition ($\omega >0$) singular and edge
branch lines are represented in Fig. 1 by solid and dashed lines,
respectively. For simplicity, the figure does not represent the
$\omega >E_u$ UHB region. (Below we find that the shape of the UHB
singular branch lines is fully determined by the shape of the
electron-removal singular branch lines represented in Fig. 1.) The
dashed-dotted lines and some of the branch lines of Fig. 1 are
border lines for the $\omega <E_u$ domain of the
$(k,\,\omega)$-plane whose spectral weight is generated by
dominant processes. (There is a region limited above by the $s$
line for $k_F<k<3k_F$ and below by the $c''$ and $c'$ lines for
$k_F<k<2k_F$ and $2k_F<k<3k_F$, respectively, which does not
belong to that domain.) The dominant processes also include
particle-hole pseudofermion processes which lead to spectral
weight both inside and outside but in the close vicinity of that
domain.

\begin{figure}
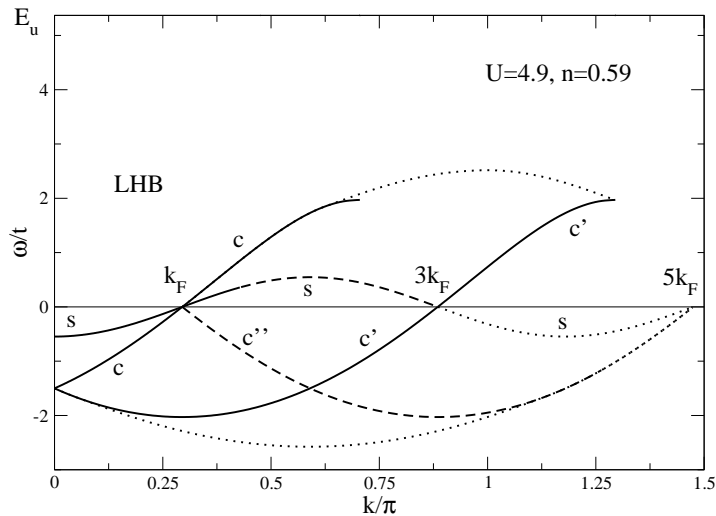

\onefigure[scale=0.4]{Fig1epl.eps} \caption{\label{fig1}
Extended-zone scheme centered at $k=0$ for $U=4.9\,t$ and
$n=0.59$. The solid and dashed lines denoted by the letters $c$,
$c'$, $c''$, and $s$ are singular and edge branch lines,
respectively. Electron removal (LHB addition) corresponds to
$\omega <0$ (and $\omega >0$) and $\omega =E_u$ marks the UHB
lower limit.}
\end{figure}

For simplicity, in this letter we consider the exponents that
control the weight shape in the vicinity of singular branch lines
only. However, similar expressions can be evaluated for any branch
line. The singular branch lines correspond to all the
$(k,\,\omega)$-plane regions where there are weight-distribution
singularities. While it is difficult to measure the exponents
experimentally, a crucial test for the suitability of the model to
describe real quasi-1D materials is whether the ARPES peak
dispersions correspond to the predicted singular branch lines. The
general branch-line spectral-function expression given in Ref.
\cite{theory} applies provided that the excited states associated
with the one-electron branch lines are used.

We start by considering the spin $s$ branch line for $0<k<3k_F$,
the charge $c$ branch line for $0<k<\pi-k_F$, and the charge $c'$
branch line for $0<k<\pi+k_F$ (see Fig. 1). The parametric
equations that define these branch lines read $\omega (k) =
\epsilon_s (q)$ for the $s$ line where $q= q (k) = (1+l)\,k_F -
l\,k$ for $(1+l)k_F/2<k<k_F + (1+l)k_F$ and $\omega (k) =
\epsilon_{\alpha} (q)$ for the $\alpha =c$ line ($\iota'=+1$) and
$\alpha =c'$ line ($\iota'=-1$) where $q = q (k) = k +\iota'\,k_F$
for $0<k< \pi -\iota'\,k_F$. Here, $l=\pm 1$ and $\epsilon_{c'}
(q)\equiv\epsilon_{c} (q)$. The following expression describes the
weight distribution in the vicinity of the $\alpha =s,c,c'$ branch
lines for $\omega$ values such that $(\epsilon_{\alpha} (q)
+l\,\omega)$ is small and positive,

\begin{equation}
B^{l} (k ,\,\omega) = C_{\alpha}^{l} (k) \Bigl(\epsilon_{\alpha}
(q) +l\,\omega\Bigl)^{\zeta_{\alpha} (k)} \, ; \hspace{0.3cm}
\alpha =s,c,c' \, . \label{BA}
\end{equation}
The $k$-dependence of $C_{\alpha}^{l} (k)$, such that
$C_{\alpha}^{l} (k)>0$ for $U/t>0$, is in general involved and can
be studied numerically. For $U/t\rightarrow 0$, $C_{\alpha}^{l}
(k)$ behaves as $C_{s}^{l}(k)\rightarrow \delta_{l,-1}\,C_{s}$,
$C^l_{c}(k)\rightarrow\delta_{l,\,+1}\,\,C_{c}$, and
$C^l_{c'}(k)\rightarrow 0$, where $C_s$ and $C_c$ are independent
of $k$. The exponent of Eq. (\ref{BA}) reads,

\begin{eqnarray}
\zeta_{s} (k) & = & - 1 + \sum_{\iota =\pm 1}\Bigl[\Bigl\{{1\over
2\sqrt{2}} - \iota\,\Phi_{s\,s}(\iota\,k_F,\,q)\Bigr\}^2  +
\Bigl\{{\iota\over 2\xi_0} +
{(1+l)\,\xi_0\over 4} - l\,\Phi_{c\,s}(\iota\,2k_F,\,q)\Bigr\}^2\Bigr]
\, ; \nonumber \\
\zeta_{\alpha}(k) & = & -1 + \sum_{\iota =\pm
1}\Bigl[\Bigl\{{f_l(\iota\,\iota')\over 2\sqrt{2}} +
\iota\,\Phi_{s\,c}(\iota\,k_F,\,q)\Bigr\}^2 + \Bigl\{{f_l(
-\xi_0)\over 4}
-\iota'\,\Phi_{c\,c}(\iota\,2k_F,\,q)\Bigr\}^2\Bigr] \, ,
\label{zetasc}
\end{eqnarray}
where $\alpha =c$ for $\iota'=+1$, $\alpha =c'$ for $\iota'=-1$,
and $f_l(x)=1-l\,(1 +x)$. In equation (\ref{zetasc}) the phase
shifts are defined in Ref. \cite{IIIb} and
$\xi_0\equiv\sqrt{2K_{\rho}}$ where $K_{\rho}$, such that
$K_{\rho}\rightarrow 1$ as $U/t\rightarrow 0$ and
$K_{\rho}\rightarrow 1/2$ as $U/t\rightarrow\infty$, is defined in
Ref. \cite{Schulz}. The exponents $\zeta_s (k)$ and
$\zeta_{\alpha=c,c'} (k)$ are plotted in Figs. 2 and 3,
respectively, as a function of $k$ for several values of $U/t$ and
$n=0.59$. The exponent $\zeta_c (k)$ (and $\zeta_{c'} (k)$) is
plotted in Fig. 3 (a) and (b) (and Fig. 3 (c) and (d)) for $l=-1$
and $l=+1$, respectively. As $U/t\rightarrow\infty$ these
exponents behave as $\zeta_{s} (k) =-1/2 + 2(q/4k_F)^2$ and
$\zeta_{s} (k) =-(q/2k_F)[1 -(q/4k_F)]$ for $l=-1$ and $l=+1$,
respectively, and $\zeta_{c}(q)=\zeta_{c'}(q) =-3/8$, in agreement
with Ref. \cite{Penc}. The weight shape of the UHB singular branch
lines is controlled by the same exponents as the corresponding
electron-removal branch lines. There is a UHB $s_u$ branch line
for $\pi-k_F<k<\pi$, $c_u$ branch line for $\pi-k_F<k<\pi$, and
${c'}_u$ branch line for $\pi-3k_F<k<\pi$. For $0<k<\pi$ and
$\omega >E_u$ the parametric equations of these lines is the same
as for the corresponding $s$, $c$, and $c'$ branch lines,
respectively, provided that $k$ is replaced by $\pi -k$ and
$\omega$ by $E_u -\omega$ in the parametric equations of the
latter lines. Under such a replacement the exponents that control
the weight shape of these UHB lines are the same as for the
corresponding electron-removal lines. While in the limit
$n\rightarrow 1$ such a correspondence refers also to the value of
the constant $C_{\alpha}^{-1} (k)$ of Eq. (\ref{BA}), otherwise
the corresponding UHB constant $C_{\alpha_u}^{+1} (\pi -k)$ is
slightly smaller. This is because for $n<1$ there is also weight
in the vicinity of four UHB $c1$ pseudofermion branch lines. We
omit here the study of the weight shape of these lines whose
exponents are positive.

\begin{figure}
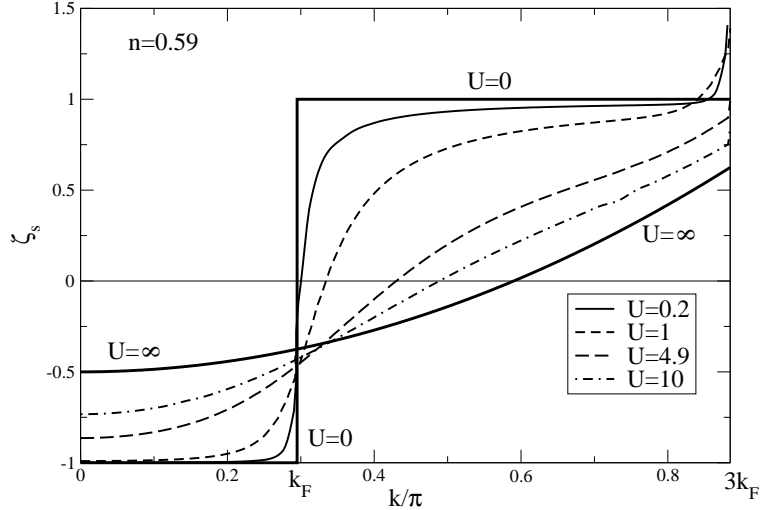

\onefigure[scale=0.4]{Fig2epl.eps} \caption{\label{fig2} Momentum
dependence along the spin $s$ branch line of the exponent
(\ref{zetasc}).}
\end{figure}

\begin{figure}
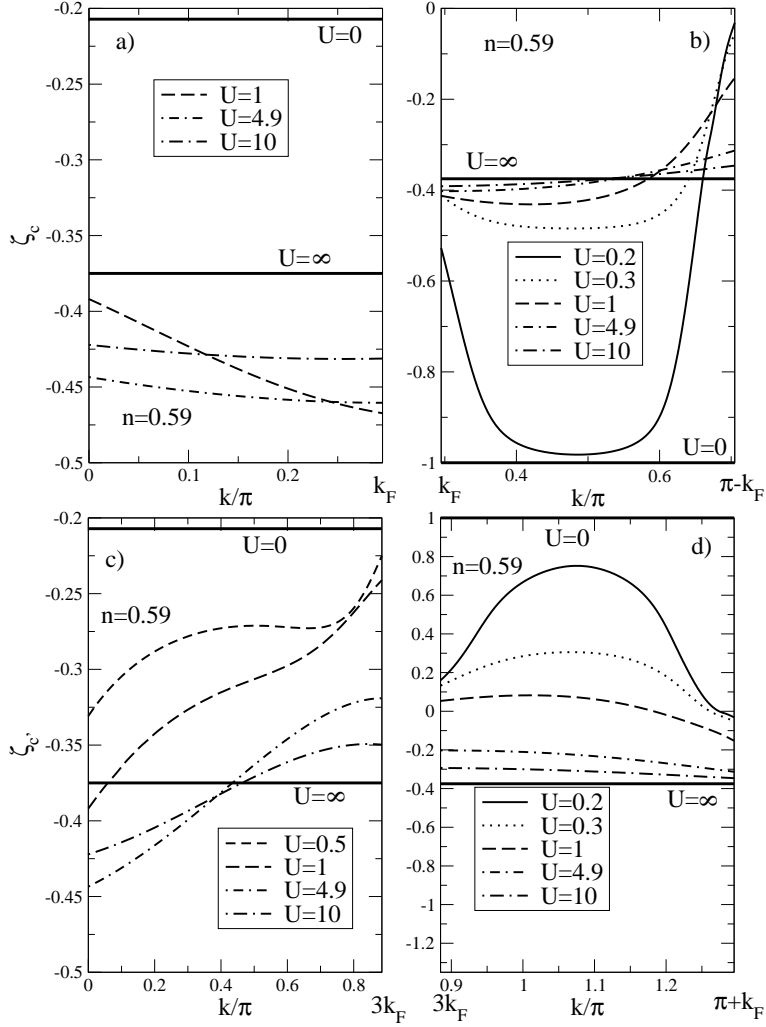

\onefigure[scale=0.4]{Fig3abepl.eps}
\onefigure[scale=0.4]{Fig3cdepl.eps} \caption{\label{fig3}
Momentum dependence of the exponent (\ref{zetasc}) along (a) and
(c) the electron-removal and (b) and (d) electron-addition charge
$c$ and $c'$ branch lines, respectively.}
\end{figure}

When the exponents (\ref{zetasc}) tend to $-1$ as $U/t\rightarrow
0$ the expression (\ref{BA}) is replaced by $\delta
(\epsilon_{\alpha} (q)+l\,\omega)$. For $0<k<\pi$ and
$U/t\rightarrow 0$ all spectral weight is transferred over to the
$s$ branch line for $0<k<k_F$, $c$ branch line for $k_F<k<\pi
-k_F$, and $s_u$ branch line for $\pi -k_F<k<\pi$. There is no
weight in other branch lines and $k>0$ regions of the
$(k,\,\omega)$-plane in such a limit. The two points
$(k=\pi-k_F,\,\omega =E_u)$ and $(k=\pi -k_F,\,\omega =\epsilon_c
(\pi))$ where $q=\pi$ corresponds to $k=\pi-k_F$ become the same
point as $U/t\rightarrow 0$ and the $c$ and $s_u$ branch lines
become connected at that point. By use of the $U/t\rightarrow 0$
expressions given in Ref. \cite{II} for $\epsilon_{c} (q)$ and
$\epsilon_{s} (q)$, one finds that these three branch lines give
rise to the electronic spectrum $\omega (k)=-2t[\cos (k)-\cos
(k_F)]$. Consistently, the above corresponding exponents are such
that $\zeta_{s} (k)\rightarrow -1$ for $0<k<k_F$, $\zeta_{c}
(k)\rightarrow -1$ for $k_F<k<\pi -k_F$, and $\zeta_{s_u}
(k)=\zeta_{s} (\pi-k)\rightarrow -1$ for $\pi -k_F<k<\pi$ as
$U/t\rightarrow 0$. Then the correct non-interacting electronic
spectral function is reached as $U/t\rightarrow 0$. (For
$U/t\rightarrow 0$ the exponents (\ref{zetasc}) behave as
$\zeta_{s} (k) = l=\pm 1$, $\zeta_{\alpha}(q) = -1/\sqrt{2} + 1/2$
for $\alpha =c,c'$ and $l=-1$, and $\zeta_{c}(q)=-1$ and
$\zeta_{c'}(q)=1$ for $l=+1$.)

We note that for $(k,\,\omega)$ points in the vicinity of the
branch-line end points $(k_0,\,j E_u)$ where $j=0$ for $\alpha
=s,c,c'$ and $j=1$ for $\alpha =s_u,c_u,{c'}_u$, respectively, the
spectral-function expression (\ref{BA}) remains valid provided
that the exponent $\zeta_{\alpha} (k)$ given in Eq. (\ref{zetasc})
is replaced by $\zeta^*_{\alpha} (k) = \zeta_{\alpha} (k) -
2\Delta_{\alpha}^{\mp 1} (k)$ for $\omega\approx
-l\epsilon_{\alpha} (q(k))=\pm l v_{\alpha} (k-k_0) + jE_u$. Here
$l=\pm 1$, $v_{\alpha}=\partial\epsilon_{\alpha}(q)/\partial
q\vert_{q=q_{F\alpha}}$, $q_{Fc}=2k_F$, $q_{Fs}=k_F$, and
$2\Delta_{\alpha}^{\mp 1} (k)$ is a functional defined in Ref.
\cite{theory}. (In Fig. 1, $k_0 =k_F$, $k_0=3k_F$, and $k_0=5k_F$
for $j=0$.) For $j =0$ this limit corresponds to the so called
low-energy Tomanaga-Luttinger liquid regime and the above exponent
$\zeta^*_{\alpha} (k)$ equals that given in Eq. (5.7) of Ref.
\cite{CFT}. The branch-line $k$ domain corresponding to that
exponent has zero measure relative to the $k$ domain of the
branch-line exponent $\zeta_{\alpha} (k)$. Therefore, in this
paper we do not consider such a limiting regime. Another case of
the interest is the behavior of the spectral function in the
vicinity of the end points $(k_0,\,j E_u)$ when these points are
approached by lines which do not cross the $\alpha$ branch lines.
In this case the weight-distribution is controlled by an exponent
different from both $\zeta_{\alpha} (k)$ and $\zeta^*_{\alpha}
(k)$ \cite{Martelo}.

\begin{figure}
\onefigure[scale=0.4]{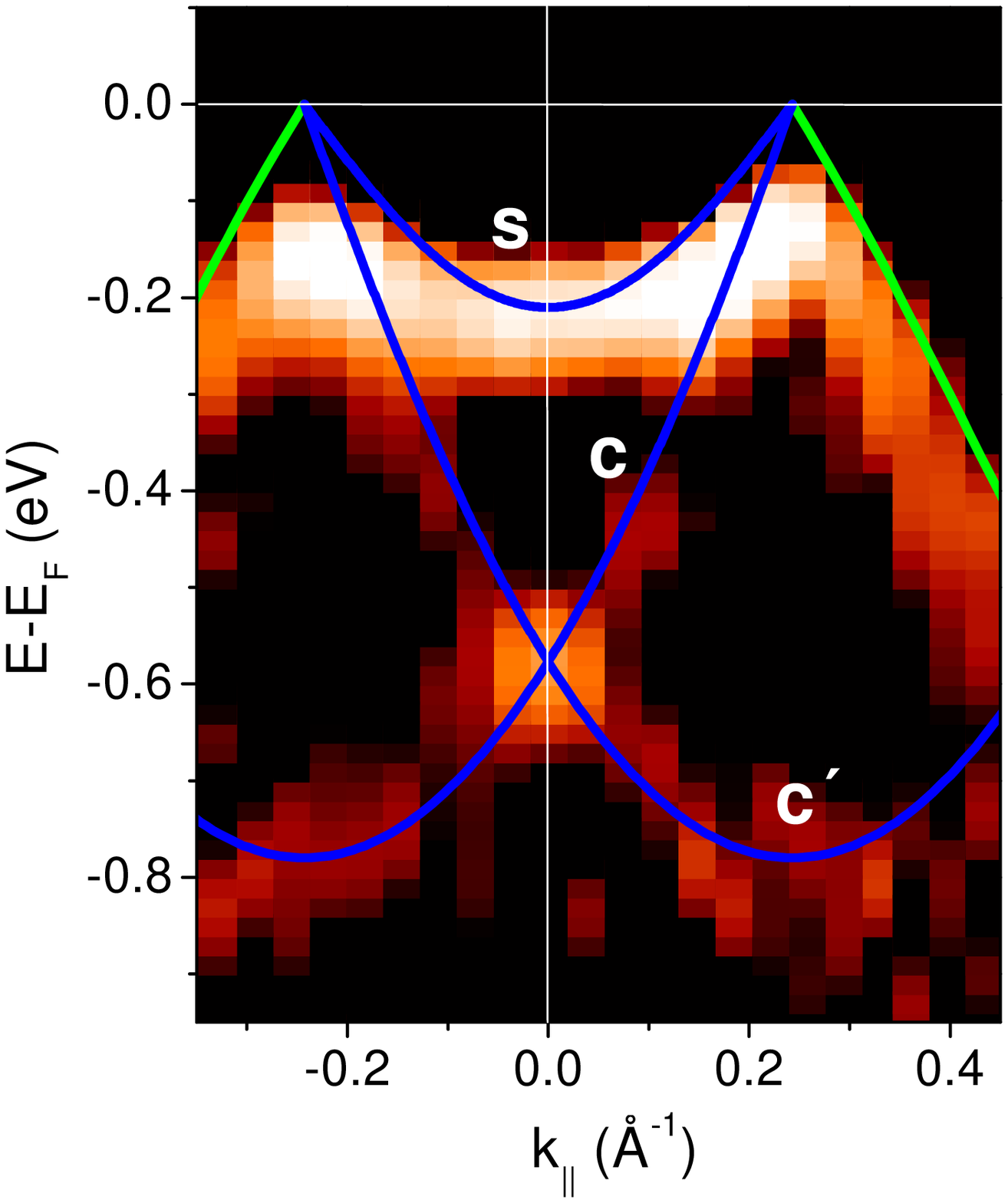} \caption{\label{fig4}
Angle-resolved photoemission spectra of TTF-TCNQ measured along
the easy transport axis and matching theoretical branch lines.}
\end{figure}

An interesting realization of a 1D metal is the organic
charge-transfer salt TTF-TCNQ \cite{Zwick98}. The experimental
dispersions in the electron removal spectrum of this quasi-1D
conductor as measured by ARPES are shown in Fig. 4. The data were
taken with He I radiation (21.2 eV) at a sample temperature of 60
K on a clean surface obtained by {\it in situ} cleavage of a
single crystal. Instrumental energy and momentum resolution amount
to 70 meV and 0.07 \AA$^{-1}$, respectively. The experimental data
in Fig. 4 reproduce earlier work \cite{Ralph,Zwick98}. While our
above theoretical weight-distribution expressions refer to all
values of $U/t$ and $n$, we find that the electron removal spectra
calculated for $t = 0.4$ eV, $U = 1.96$ eV ($U/t = 0.49$), and
$n=0.59$ yields an almost perfect agreement with the three TCNQ
experimental dispersions. The exception is the low-energy
behavior, as a result of the inter-chain hopping and
electron-phonon interactions, as mentioned above. If accounted for
a renormalization of the transfer integral due to a possible
surface relaxation \cite{Ralph}, these values are in good
agreement with estimates from other experiments \cite{Zwick98}.
The experimental TCNQ finite-energy peak dispersions of Fig. 4
correspond to the spin $s$ branch line and charge $c$ and $c'$
branch lines. Importantly, only these singular branch lines, whose
weight shape is controlled by negative exponents, lead to peak
dispersions in the real experiment. The other peak dispersion of
Fig. 4 is associated with the electron-removal spectrum of the TTF
chains whose density is $n=2-0.59=1.41$. It can be described by
the electron-addition spectrum of a corresponding $n=0.59$
problem. Remarkably, we find again that the theoretical lines
match the TTF dispersion provided that $t\approx 0.27$ eV and $U<
0.2t$ within experimental uncertainty. Indeed, for such small
values of $U/t$ and $n=0.59$ both the electron-addition exponents
of Figs. 2 and 3 (d) are positive, whereas the exponent of Fig. 3
(b) and the UHB exponent $\zeta_{s_u} (k)=\zeta_{s} (\pi-k)$ for
$\pi-k<k_F$ (see Fig. 2) are negative. For $n=1.41$ the $c$ and
$s_u$ lines associated with these exponents correspond to electron
removal and thus appear at negative values of $\omega$ and lead to
a single singular branch line formed by the $c$ line for
$0.59\pi/2<k<1.41\pi/2$ and the $s_u$ line for $1.41\pi/2<k<\pi$.
The energy pseudogap at $k=1.41\pi/2$, which separates the $c$
line from the $s_u$ line, nearly vanishes for $U/t<0.2t$ and thus
is not observed in the experiment. This singular branch line is
that matching the TTF peak dispersion of Fig. 4. (It matches such
a dispersion for $0.59\pi/2<k<\pi$.) We thus conclude that in
contrast to the conventional band structure description, the
singular branch lines obtained from the 1D Hubbard model describe
quantitatively, for the whole finite-energy band width, the peak
dispersions observed by ARPES in TTF-TCNQ. This seems to indicate
that the dominant non-perturbative many-electron microscopic
processes described by the pseudofermion dynamical theory of Ref.
\cite{theory}, which control the weight distribution of these
charge and spin singular branch lines, also control the unusual
finite-energy one-electron spectral properties of TTF-TCNQ.

We thank E. Jeckelmann for stimulating discussions. K.P. thanks
the financial support of OTKA grants D032689 and T037451, L.M.M.
of FCT grant BD/3797/94, and R.C., M.S., and U.S. of Deutsche
Forschungsgemeinschaft (CL 124/3-3 and SFB 484).


\begin{thebibliography}{99}
\bibitem{Ralph}\Name{Sing M., Schwingenschl\"ogl U., Claessen R.,
        Blaha P., Carmelo J. M. P., Martelo L. M., Sacramento P. D.,
        Dressel M. \and Jacobsen C. S.}
        \REVIEW{Phys. Rev. B}{68}{2003}{125111}.
\bibitem{Zwick98} \Name{Claessen R., Sing M., Schwingenschl\"ogl U., Blaha P.,
        Dressel M. \and Jacobsen C. S.}
        \REVIEW{Phys. Rev. Lett.} {88}{2002}{096402}; \Name{Zwick F.,
        J\'erome D., Margaritondo G., Onellion M., Voit J. \and
        Grioni M.}\REVIEW{Phys. Rev. Lett.} {81}{1998}{2974};
        \Name{Kagoshima S., Nagasawa H. \and Sambongi T.}
        {\em One-dimensional conductors} (Springer, Berlin){1987} and
        references therein.
\bibitem{Arikawa}
        \Name{Arikawa M., Saiga Y. \and Kuramoto Y.}
        \REVIEW{Phys. Rev. Lett.}{86}{2001}{3096};
        \Name{Penc K.\and Shastry B. S.}
        \REVIEW{Phys. Rev. B}{65}{2002}{155110}.
\bibitem{Schulz}
        \Name{Schulz H. J.}
        \REVIEW{Phys. Rev. Lett.}{64}{1990}{2831}.
\bibitem{CFT}
        \Name{Frahm H. \and Korepin V. E.}
        \REVIEW{Phys. Rev. B}{43}{1991}{5653} and references therein.
\bibitem{Penc}
        \Name{Penc K., Hallberg K., Mila F. \and Shiba H.}
        \REVIEW{Phys. Rev. Lett.}{77}{1996}{1390};
        {\it ibid.}
        \REVIEW{Phys. Rev. B}{55}{1997}{15475}.
\bibitem{Sorella}
        \Name{Sorella S. \and Parola A.}
        \REVIEW{Phys. Rev. Lett.}{76}{1996}{4604}.
\bibitem{Senechal}
        \Name{S\'en\'echal D., Perez D. \and Pioro-Ladri\`ere M.}
        \REVIEW{Phys. Rev. Lett.}{84}{2000}{522}.
\bibitem{I}
        \Name{Carmelo J. M. P., Rom\'an J. M. \and Penc K.}
        \REVIEW{Nucl. Phys. B}{683}{2004}{387}.
\bibitem{II}
        \Name{Carmelo J. M. P. \and Sacramento P. D.}
        \REVIEW{Phys. Rev. B}{68}{2003}{085104}.
\bibitem{IIIb}
        \Name{Carmelo J. M. P.} cond-mat/0305568.
\bibitem{theory}
        \Name{Carmelo J. M. P. \and Penc K.}
        cond-mat/0311075.
\bibitem{Martelo}
        \Name{Carmelo J. M. P., Martelo L. M. \and Sacramento P. D.}
        \REVIEW{J. Physics: Cond. Matt.}{16}{2004}{1375}.
\end{thebibliography}
\end{document}